\RequirePackage{etoolbox}
\csdef{input@path}{%
 {sty/}
 {img/}
}%
\csgdef{bibdir}{bib/}

\documentclass[ba]{imsart}
\pubyear{0000}
\volume{00}
\issue{0}
\doi{0000}
\firstpage{1}
\lastpage{1}

\usepackage{amsthm}
\usepackage{amsmath}
\usepackage{natbib}
\usepackage[colorlinks,citecolor=blue,urlcolor=blue,filecolor=blue,backref=page]{hyperref}
\usepackage{graphicx}

\startlocaldefs
\numberwithin{equation}{section}
\theoremstyle{plain}

\endlocaldefs

\begin{document}

\begin{frontmatter}
\title{Invited Discussion of ``A Unified Framework for De-Duplication and
Population Size Estimation''}
\runtitle{Invited Discussion}

\begin{aug}
\author{\fnms{Jared S.} \snm{Murray}\thanksref{addr1,t1,t2,m1}\ead[label=e1]{jared.s.murray@mccombs.utexas.edu}}

\runauthor{J. S. Murray}

\address[addr1]{Department of Information, Risk, and Operations Management and Department of Statistical Science. University of Texas at Austin. 
    \printead{e1} 
}

\thankstext{t2}{Supported by SES-1824555
The author gratefully acknowledges support from the National Science Foundation under grant number SES-1824555. Any opinions, findings, and conclusions
or recommendations expressed in this material are those of the author(s) and do not necessarily reflect the
views of the funding agencies.
}

\end{aug}

\end{frontmatter}

I would like to congratulate the authors on a stimulating contribution to the literature on record linkage/de-duplication and population size estimation. \cite{tancredi2011hierarchical} was one of the papers that first piqued my interest in record linkage, so I am pleased to see more work along these lines (with an author population size of N+1!)  My discussion below focuses on two main themes: Providing a more nuanced picture of the costs and benefits of joint models for record linkage and the ``downstream task'' (i.e. whatever we might want to do with the linked and de-duplicated files), and how we should measure performance.

\section{The promise and peril of joint modeling: A partial defense of disunity}

The promise of a joint model for record linkage, de-duplication, and population size estimation is likely obvious to the readership of Bayesian Analysis: We immediately obtain valid posterior inference over the population size that accounts for uncertainty about duplicates and links across files -- provided that we specify an adequate joint model. Which leads us predictably to the peril of joint modeling, the fact that specifying a model for any of these three tasks alone is nontrivial. Addressing them simultaneously in a single model requires specifying a joint model sufficiently rich to do well on all three tasks (linkage, de-duplication, and population size estimation) while being tractable enough to understand its properties and perform posterior inference. 

The model presented here necessarily makes some compromises in service of joint modeling, and I wonder about their impact. For example, assumptions about the sampling process generating the lists are essential to modeling the unknown population size and therefore must appear in any unified model.  This will consequently restrict the prior distribution over the overlap between files in the record linkage/de-duplication portion of the model, despite the fact that the assumption of simple random sampling from the population -- or any sort of random sampling at all -- is otherwise irrelevant to record linkage and de-duplication.  The assumptions made by the authors imply a very particular, informative prior distribution on $Z$, the partition of records into co-referent sets, and therefore on $K$, the number of distinct units captured across all lists (as reported in Table 1).  

This choice is consequential. Indeed, immediately prior to Section 3.1 the authors note that the induced prior distribution on $K$ is probably {\em not} well-suited to record linkage tasks in general, which makes me wonder why we should expect it to work well when doing record linkage and population size estimation simultaneously. I have to assume that either 1) we actually don't expect it to work particularly well but the joint model at hand demands it or 2) the assumptions about the sampling process are actually warranted here, at least approximately, while they may not be in general applications of record linkage. If the former, this seems to beg the question and ignore options beyond joint modeling.  If the latter, things are more interesting. 

If the assumptions are in fact correct, we would expect to obtain more accurate and efficient inferences by inducing the ``true'' prior over $Z$ and $K$ using the joint model. But what happens when the sampling assumptions are violated? It is difficult to say, and it must depend on a host of factors (such as the degree and frequency of errors among co-referent records). However, it is not hard to imagine a case where relatively minor deviations from the sampling assumptions are more or less innocuous in the context of a population size model with known partition $Z$ but become influential when $Z$ is unknown and jointly modeled, due to the influence of the ``misspecified'' informative prior over $Z$.  It would be interesting to try and draw this out via a simulation exercise {particularly in light of how influential \cite{steorts2016} found a similar prior to be in a pure record linkage/de-duplication context).

If posterior inference is not robust to deviations from the sampling assumptions, what could we do instead? The desire to mitigate this undesirable ``feedback'' from a misspecified sub-model appears in many different settings, from Bayesian causal inference with propensity score models \citep{mccandless2010cutting,zigler2013model} to astrophysics \citep{yu2018incorporating} and beyond (see \citet{jacob2017better} for additional examples). This is a difficult problem and an active area of research. The proposed solutions often take the form of (possibly incoherent) multistage inference, in this case inferring the linkage structure in stage 1 and the population size in stage 2, propagating uncertainty from stage 1 to stage 2 without allowing any information from stage 2 to flow to stage 1.  \citet{jacob2017better} give examples of settings where these ``posteriors'' are better than the posterior under a misspecified joint model in a decision-theoretic sense.

In the context of de-duplication and population size modeling, \cite{sadinle2018} proposes a related two-stage alternative to joint modeling termed ``linkage averaging''. If (in the notation of the current paper) $h(\lambda)$ is the estimate of population size we would compute given complete data (i.e., a de-duplicated and linked set of files) then under certain conditions the posterior for $h(\lambda)$ under a record linkage/de-duplication model alone will give the same inferences as a proper Bayesian joint model for linkage, de-duplication, and population size estimation.  With a single set of posterior samples one can perform inference over multiple models for the population size, again provided that they all satisfy some relatively mild conditions.

These conditions do necessarily demand a degree of compatibility between the prior on $\lambda$ and the population size model. They bear a striking similarity to the conditions under which multiple imputation delivers (asymptotically) valid Bayesian inference (``congeniality'', \citep{meng1994multiple,xie2017dissecting,murray2018multiple}). This raises the interesting question of whether the compatibility conditions might be relaxed while still yielding conservative inferences, similar to the way one can obtain conservative inferences using imputations under an uncongenial imputation model, provided it is uncongenial in the ``right'' way (roughly, by making fewer assumptions during imputation than analysis).  

\section{Measuring and improving performance}

Various sub-specialties of statistics have spawned their own de-facto benchmark datasets -- think of the iris data for clustering or the galaxy dataset for density estimation. Likewise, \texttt{RLdata500} and \texttt{RLdata10000} have arguably become something of a benchmark in record linkage problems due in large part to their accessibility via the popular \texttt{RecordLinkage} R package. I have used them in publications myself \citep{Murray_2015}. Benchmark datasets form a sort of lingua franca that is useful for teaching, exposition, and as a sort of sanity check (when our brilliant new method finds six distinct clusters in the iris data, it's back to the drawing board). 

However, we have to be careful extrapolating from these datasets to more complex settings. In the provocatively titled ``Leave the Pima Indians Alone'', \citet{chopin2017leave} make the case that an excessive focus on relatively simple binary regression problems like the Pima Indians diabetes dataset has had a distortive impact on the Bayesian computation literature. I worry a little that repeatedly going back to the \texttt{RLdata} datasets might lead the record linkage literature up the same path. In particular, the errors in these synthetic datasets are rather minimal, and the duplicate record pairs are quite well-separated from the non-duplicates.  In my experience this not representative of the datasets we see in the wild, at least not those that demand sophisticated statistical modeling. 
Like Britney and the Pima Indians, I think it may be time to leave \texttt{RLdata} alone.  

However, the primary evidence that the authors provide in favor of their model is its performance on \texttt{RLdata} datasets. Even setting aside whether this is a representative testbed, I wonder if this is much evidence at all since no alternative approaches are presented. Several are available, at least for the record linkage and de-duplication tasks, including some developed by the authors themselves (e.g. \cite{steorts2015} reports FNR and FDR of 0.02 and 0.04 on \texttt{RLdata500}, versus 0.015 and 0.08 using the model in the current paper).  How well do existing Bayesian models perform on the linkage/de-duplication task? What about even simpler methods, like the point estimates generated by Fellegi-Sunter methods \citep{fellegi1969theory} or their generalizations \citep{sadinle2013generalized,Murray_2015}? This is important context; while the model proposed here offers richer inference, should we trust those inferences if the model does not perform relatively well on the linkage/de-duplication task?

The authors actually seem to go a step further and use results on {\texttt RLdata} to inform parameter selection when modeling the Syrian casualty data. This frankly seems like a bad idea; in my own experience with similar files \citep{syria}, including expert hand-linked datasets, we observed very different patterns of distortion among co-referent records than the simple patterns one would find in {\texttt RLdata}.  Given how variable performance is across parameter settings in Section 4, I would suggest that at least some sensitivity analysis might be in order for the Syria application.

Rather than rely on unrepresentative benchmark datasets to measure performance and select parameters, what could we do instead? The longer I work on record linkage problems the more I am convinced of the need to include a hand-labeling exercise alongside every serious application. The synthetic datasets at our disposal are limited in the range of errors they include and are often poor representations of the problem at hand. Model-based estimates of error rates are only as good as the model, and if we're not sure about the model... However, provided that the true error rates are low, precise estimates of false match rates (false discovery rates) can be obtained via random sampling from matched record pairs. False match rates aren't everything, but they aren't nothing either. Sadly the authors missed an opportunity to do even a little inspection here; after finding a small number of duplicates in the Syria application, they note only that ``visual inspection of these pairs may eventually confirm their matching status''.

Ideally a labeling exercise to evaluate a record linkage/de-duplication method should include matches generated by other methods (to remove potential bias toward declaring estimated matches correct), blinding (to the method(s) that declared the link), multiple review, an ``indeterminate'' or ``unsure'' option for the labelers, and should present labelers with neighboring ``near-match''record pairs. Stellar examples of hand-labeling study designs include \cite{bailey2017well,8637549}. In \cite{posthoc} we hand-labeled a relatively small number of links to compare two competing methods, including one Bayesian model. For the Bayesian model we also used these labels to obtain the posterior distribution of false match rate adjusted estimators by computing them on each posterior sample of the linkage structure (similar to \cite{sadinle2018}'s linkage averaging). For our estimands, we only found it necessary to adjust for the false match rate and we did not grapple with simultaneous de-duplication or multiple files.  But we did find that variation due to assumptions about bias from linkage error tended to swamp variation due to uncertainty about the linkage structure. 

Reducing or otherwise accounting for linkage error seems important in the context of the current paper as well. Observe that in Figure 3, the estimates of $K$ are worse in the blocks with higher error rates (blocks 7, 1, 10, 3) and in each case the estimate for $K$ is biased down with a rather concentrated posterior distribution. If the model cannot be improved further, perhaps we would be better off looking at the posterior distribution of linkage error adjusted estimates of the population size.  Linkage error adjusted estimators for the population size do exist, at least for relatively simple settings (e.g. \citet{ding1994dual,di2018population,soton436665}) and perhaps could be cast in \cite{sadinle2018}'s framework of linkage averaging (although I have not checked the compatibility conditions myself).  These estimators depend on false non-match rates, which are more difficult to obtain through hand labeling but often can be reasoned about based on plausible levels of duplication and overlap.  This reasoning could form the basis of a computationally efficient sensitivity analysis. This seems like a promising avenue for future research, alongside further improvements in model and prior specification to minimize error rates.



\bibliographystyle{ba}
\bibliography{sample}


\end{document}